\documentclass[twocolumn,showpacs,preprintnumbers,amsmath,amssymb,APSl,prd,nofootinbib,superscriptaddress]{revtex4-2}

\usepackage{bm}
\usepackage{mathrsfs}
\usepackage{xcolor,color,graphicx,graphics}
\usepackage{epsfig,subfigure}
\usepackage{latexsym,amssymb,amsmath,amsfonts} \usepackage[english]{babel} \usepackage[OT1]{fontenc}
\usepackage[latin1]{inputenc}
\usepackage{makeidx}
\usepackage{hyperref}
\usepackage{color,graphicx,graphics,wrapfig,epsf}\usepackage{hyperref}\hypersetup{colorlinks=true, linkcolor=blue, citecolor=green}
\usepackage{makecell}
\usepackage{multirow}
\usepackage{physics}

\usepackage{tikz}
\usetikzlibrary{decorations.pathmorphing, arrows.meta, calc, intersections}

\usepackage{appendix}
\usepackage{orcidlink}

\usepackage{lipsum}     
\usepackage{mathtools}

\usepackage{latexsym}
\usepackage{orcidlink}
\usepackage{enumerate}

\definecolor{responseblue}{RGB}{0,70,160}

\definecolor{fixorange}{RGB}{255,128,0}

\definecolor{red}{rgb}{1,0,0}

\def\+{^\dagger}

\def\<{\leftarrow}
\def\>{\rightarrow}

\def\({\left(}
\def\){\right)}

\newcommand{\der}{{d}}
\newcommand{\bi}{\begin{itemize}} 				\newcommand{\ei}{\end{itemize}}
\newcommand{\benu}{\begin{enumerate}} 		\newcommand{\enu}{\end{enumerate}}
\newcommand{\bd}{\begin{dinglist}{0}}     \newcommand{\ed}{\end{dinglist}}
\newcommand{\bfig}{\begin{figure}[htbp]}  \newcommand{\efig}{\end{figure}}
        			
\newcommand{\bc}{\begin{center}} 				  \newcommand{\ec}{\end{center}}
\newcommand{\be}{\begin{equation}} 				\newcommand{\ee}{\end{equation}}
\newcommand{\bsub}{\begin{subequations}}  \newcommand{\esub}{\end{subequations}}
\newcommand{\ben}{\begin{eqnarray}} 			\newcommand{\een}{\end{eqnarray}}
\newcommand{\ba}[1]{\begin{array}{#1}} 		\newcommand{\ea}{\end{array}}
\newcommand{\bea}{\begin{equation}\begin{array}{rcl}}
\newcommand{\eea}{\end{array}\end{equation}}

\usepackage{soul}

\begin{document}
\title{Photon spheres in dynamical spacetimes}

\author{David D\'iaz-Guerra
\orcidlink{0000-0003-0859-917X}} 
\email{ddiazgue@ucm.es}
\affiliation{Departamento de F\'isica Te\'orica, Universidad Complutense de Madrid, E-28040 Madrid, Spain}

\author{\'Angel Rinc{\'o}n 
        \orcidlink{0000-0001-8069-9162}
        }        \email{angel.rincon@physics.slu.cz}
 \affiliation{Departamento de F{\'i}sica, Universidad del B{\'i}o-B{\'i}o,
Casilla 5-C, Concepci{\'o}n, Chile.}
        \affiliation{Research Centre for Theoretical Physics and Astrophysics, Institute of Physics, Silesian University in Opava, Bezrucovo n\'{a}m. 13, 74601 Opava, Czech Republic.}

\author{Diego Rubiera-Garcia 
\orcidlink{0000-0003-3984-9864}
}
\email{drubiera@ucm.es}
\affiliation{Departamento de F\'isica Te\'orica and IPARCOS, Universidad Complutense de Madrid, E-28040 Madrid, Spain}

\date{\today}
\begin{abstract}
    The characterization of the photon region, i.e. the region of space-time filled with unstable bound null geodesics, is critical to understand the behavior of radiation near a compact-enough object, such as black holes. However, its study has been focused on stationary space-times, neglecting dynamical scenarios such as collapsing, accreting, or evaporating black holes. 
        In this work, we present a novel covariant approach to describe radiation in dynamical spherical space-times. Using a $2+2$ decomposition, we characterize a dynamical photon sphere as the two-sphere associated with instantaneous critical turning points of non-radial null motion, where both the radial momentum and its derivative along the null ray vanish. A smooth time-dependent family of these spheres defines a photon sphere tube, which is not assumed to be a photon surface generated by null geodesics. The resulting condition is quasi-local and can be written algebraically in terms of the areal radius, the Misner-Sharp mass, and the radial pressure. We further use geodesic deviation to characterize the local radial deviation between neighboring null geodesics at each critical turning point, while distinguishing this local deviation from the global stability of an evolving null orbit. Our formulation recovers the standard photon sphere in the static limit, and we show dynamical examples like stellar collapse and accreting/evaporating scenarios.
\end{abstract}

\maketitle

\section{Introduction}

The photon region, namely, the set of space-time points admitting (typically unstable) bound null geodesics, is fundamental to the physics of ultra-compact objects \cite{Chandrasekhar:1985kt,Teo:2003ltt}, which includes, but it is not limited to, black holes. This is so because it plays a key role in determining the optical appearance of such objects, by defining the region where photons are (even if temporarily) trapped, and underlying the development of both the black hole shadow and the bright region surrounding it \cite{Cunha:2018acu}. Furthermore, during the ringdown phase of binary mergers, gravitational wave (GW) perturbations are dominated by quasi-normal modes, whose properties are closely related to the (unstable) photon region and which govern their propagation and decay \cite{Cardoso:2008bp}. Furthermore, horizonless ultra-compact objects generically posses a pair of such surfaces, one unstable and one stable \cite{Cunha:2017qtt}, such that the latter may support long-lived trapped modes which potentially lead to non-linear instabilities that destabilize the system \cite{Cardoso:2014sna,Cunha:2022gde}, threatening the astrophysical viability of any such objects. Therefore, the precise characterization of photon surfaces provides a crucial link between gravitational theory and observational phenomenology \cite{Cardoso:2016rao}. This is of particular relevance after the finding, within the last decade, of astronomical events suitably identified as corresponding to each optical/GW scenario, as demonstrated by the Event Horizon Telescope observations of M87$^*$ \cite{EventHorizonTelescope:2019dse} and Sgr A$^*$ \cite{EventHorizonTelescope:2022wkp}, and by those of the LIGO-VIRGO-KAGRA Collaborations of dozens of binary black hole mergers \cite{LIGOScientific:2025slb}.

The standard conceptual and technical framework of the photon region is formulated for stationary space-times, which is sufficient in regimes where the time variations in the black hole geometry are slow as compared to the relevant astrophysical time-scale variations, so the latter can be  safely neglected \cite{Cunha:2020azh}. However, this approximation fails to account for both theoretical and phenomenological scenarios of interest, such as gravitational collapse, fully-dynamical accretion processes, or black hole evaporation. Furthermore, it prevents us from rigorously studying objects in which there is a time-dependence on the metric functions defining the space-time geometry, as happens, for instance, in certain classes of bosonic star configurations, oscillations, or Q-balls, see e.g.  \cite{Friedberg:1986tq,Seidel:1991zh,Alcubierre:2003sx,Liebling:2012fv,Visinelli:2021uve,Herdeiro:2021lwl,deSa:2025nsx,Macedo:2025wey,Jaramillo:2026ygy}.

In dynamical space-times, the photon region can be characterized in terms of photon surfaces: hypersurfaces for which initially tangent null geodesics remain tangent~\cite{Claudel:2000yi}. In particular space-times, a critical photon surface can be selected by integrating the null-geodesic equations with a stationary photon surface condition~\cite{Mishra:2019trb,Solanki:2022glc,Giambo:2025jru,Koga:2022dsu}. By tracking null orbits over an extended region, this approach shows global trapping properties tied to the chosen boundary condition. By contrast, quasi-local approaches are pointwise definitions, including constructions based on codimension-two geometry~\cite{Cao:2019vlu} and dynamically transversely trapping surfaces (DTTSs)~\cite{Yoshino:2019dty}. Recent numerical calculations also identify dynamical light rings on individual time slices through instantaneous turning-point conditions~\cite{Zhang:2025lhm}.

Following this quasi-local perspective, we introduce a covariant definition of a \textit{dynamical photon sphere} as the two-sphere selected by instantaneous critical turning-point conditions on non-radial null motion. Using a $2+2$ decomposition and the normalized Kodama-radial frame, these conditions yield an algebraic equation for the photon sphere radius in terms of the Misner-Sharp mass and the radial pressure, extending the static limit without integrating the orbit or imposing global boundary conditions.

This dynamical framework reveals important insights. For instance, we demonstrate that dynamical photon spheres can form independently of trapping horizons, such as within a collapsing star before it becomes a black hole. We also characterize the local radial deviation between neighboring null geodesics at each critical turning point in the time-dependent family of dynamical photon spheres.
While observations related to dynamical photon regions are still beyond our current observational reach, and despite its relatively smaller impact on the interpretation of astronomical data,  this problem is of interest both theoretically and phenomenologically, which is proven by the exploration of dynamical compact objects in the last few years as discussed above.

This paper is organized as follows. In Sec.~\ref{sec:formalism}, we establish the covariant formalism for null geodesics in spherical space-times. In Sec.~\ref{sec:photon_surfaces}, we define dynamical photon spheres through instantaneous critical turning-point conditions and derive their quasi-local location and radial perturbation equations. In Sec.~\ref{sec:examples}, we apply the formalism to static geometries, stellar collapse, and accreting or evaporating space-times. We conclude in Sec.~\ref{sec:conclusion}. An appendix is included in Sec.~\ref{sec:gravitational_field_equations} to summarize the gravitational field equations.

\section{Null geodesics in dynamical spherical space-times} \label{sec:formalism}

\subsection{2+2 spherical decomposition}

We consider a dynamical and spherically symmetric space-time $\mathcal{M}$, which allows for a 2+2~decomposition of the manifold into a time-radial part $\mathcal{M}_s$, with coordinates $x^a$ (lowercase Latin letters), and the unit 2-sphere $S^2$, with standard angular coordinates $(\theta, \phi)$. Following this decomposition, the metric tensor is
\begin{equation}
	g_{\mu \nu} \der x^\mu \der x^\nu = g_{ab}(x^c) \der x^a \der x^b + r^2(x^c) \der \Omega^2.
\label{eq:sph_metric_bg}
\end{equation} 
The term $g_{ab}$ is the metric of the 2-dimensional time-radial submanifold $\mathcal{M}_s$, while $\der\Omega^2 = \der\theta^2 + \sin^2\theta\der\phi^2$ is the angular metric of $S^2$. The areal radius $r(x^a)$ is a scalar field on $\mathcal{M}_s$, such that the area of a 2-sphere of the system is $4\pi r^2$.

The gradient of the areal radius $r(x^a)$ defines a vector
\begin{equation}
	r_a \equiv \partial_a r(x^b), \label{eq:def_radial}
\end{equation} 
that describes the normal direction to the surfaces of the 2-sphere. We define the functions $f(x^a)$ and $m(x^a)$, and relate them as
\begin{equation}
	f(x^a) \equiv r_a r_b  g^{ab}= 1 - \frac{2m(x^a)}{r(x^a)}. \label{eq:def_f}
\end{equation}
The function $m(x^a)$ is the Misner-Sharp mass~\cite{Misner:1964je,Hernandez:1966zia}, which represents the active gravitational energy of the system. 

A time-like Killing vector, representing a time-translation symmetry, is not guaranteed to exist in a general dynamical space-time. However, for any spherically symmetric metric, there exists the Kodama vector \cite{Kodama:1979vn}. It is defined on the time-radial manifold $\mathcal{M}_s$ as
\begin{equation}
    t^a = -\varepsilon^{ab} r_b, \label{eq:def_kodama}
\end{equation}
where $\varepsilon^{ab}$ is the volume form on $\mathcal{M}_s$. By construction, the Kodama vector is orthogonal to the radial vector, $t^a r^a = 0$, and its norm is given by $t^a t^b g_{ab} = -f(x^a)$. For any stationary space-time, the Kodama vector is proportional to the time-like Killing vector related to the time symmetry.

The existence of both the Kodama~\eqref{eq:def_kodama} and the radial vectors~\eqref{eq:def_radial}, allows us to build a vector basis on the time-radial manifold. Instead of the Kodama and radial vectors, we use their normalized counterparts, $\{u_a, n_a\}$, defined as
\begin{equation}
    u_a = \frac{t_a}{\sqrt{f}}, \quad  n_a = \frac{r_a}{\sqrt{f}},
\end{equation}
such that
\begin{align}
    u_a u^a &= -1, & n_a n^a &= 1, & u_a n^a &= 0.
\end{align}

We can next perform a kinematic decomposition on the vector basis (here a semicolon denotes the covariant derivative on the $\mathcal{M}_s$ time-radial manifold) as
\begin{align}
    u_{a:b} &= n_a \left( \theta_u n_b - \theta_n  u_b \right). \\
    n_{a:b} &= u_a \left( \theta_u n_b - \theta_n  u_b \right),
\end{align}
with the scalar expansions of the vectors defined as
\begin{align}
    \theta_u &\equiv \nabla^a u_a = - D_u \ln \sqrt{f}, \\
    \theta_n & \equiv \nabla^a n_a = \frac{\square r}{\sqrt{f}}- D_n \ln \sqrt{f}.
\end{align} 
where we have expressed the time-radial d'Alembertian as $\square \equiv g^{ab}\nabla_a \nabla_b$, and the directional derivatives of a scalar $\phi$ along the $u_a$ and $n_a$ vectors as
\begin{align}
    D_u \phi &\equiv u^a \partial_a \phi, & D_n \phi &\equiv n^a \partial_a \phi.
\end{align}

One key aspect to address is that the causal structure of the regions of the space-time is determined by the sign of $f(x^a)$\cite{Hernandez:1966zia,Hayward:1993wb}, as it determines the causal character of the basis vectors. In untrapped (trapped) regions, the radial vector, $r_a$, is space-like (time-like) and the Kodama vector, $t_a$, is time-like (space-like). The boundary between these regions, where $f=0$, defines the location of trapping horizons $x_\text{h}$ as
\begin{equation}
    f(x_\text{h}) = 0 \implies \left[ r - 2m \right]_{x_\text{h}} = 0. \label{eq:trap_surface}
\end{equation}
A rigorous study on trapping horizons can be found in Ref.~\cite{Nielsen:2005af}. On these trapping surfaces, both the radial and Kodama vectors are null. An equivalent argument on the causality follows on their normalized versions $u_a$ and $n_a$. However, these vectors are not defined on the trapping horizon.

Therefore, the time-like vector $u_a$ and the space-like $n_a$ form an orthonormal basis on the two-dimensional time-radial manifold $\mathcal{M}_s$, as long as we do not pass through a trapping horizon. We particularize our analysis for the untrapped regions which are physically reachable without going through a horizon.

\subsection{Null geodesics in dynamical spherical space-times}
We describe a geodesic by its tangent vector $k^\mu = \frac{\der x^\mu}{\der s}$, with $s$ the parameter labeling the curve. We focus on null geodesics that satisfy the null equation:
\begin{equation}
    g_{\mu\nu} k^\mu k^\nu = 0. \label{eq:null_norm}
\end{equation}
Due to spherical symmetry, the motion can be separated into a time-radial part and an angular part, that is
\begin{equation*}
    k_\mu = (k_a, k_A).
\end{equation*}
Then, the null equation \eqref{eq:null_norm} can be written as
\begin{equation}
    g^{ab} k_a k_b + \frac{L^2}{r^2} =  0, \label{eq:null_geodesic}
\end{equation} 
where $L^2$ is the squared norm of the angular part of the geodesic, and corresponds to the conserved squared angular momentum of the geodesic, as a consequence of the spherical symmetry of the space-time.

The evolution of the tangent vector along the geodesic is described by the geodesic equation. For an affinely parameterized geodesic, this equation is:
\begin{equation}
    \nabla_k k_\mu = 0, \label{eq:geodesic_evolution}
\end{equation}
where $\nabla_k (\cdot) \equiv k^\alpha \nabla_\alpha (\cdot)$ is the directional covariant derivative along the geodesic. On the time-radial manifold $\mathcal{M}_s$, we decompose the momentum vector $k_a$ using the geometric basis provided by the normalized Kodama $u_a$ and radial $n_a$ vectors as
\begin{equation}
    k_a = \Omega u_a + \Pi n_a. \label{eq:k_decomposition}
\end{equation}
This projection defines the  local energy $\Omega(x^a)$ measured in the normalized Kodama frame and the radial momentum $\Pi(x^a)$ as the corresponding projections of the tangent vector onto the normalized Kodama and radial vectors:
\begin{align}
    \Omega &\equiv - k_a u^a, \label{eq:k_Omega} \\
    \Pi    &\equiv k_a n^a \ , \label{eq:k_Pi}
\end{align}
respectively. The decomposition provides a covariant approach to null geodesics in a general dynamical setup. However, the decomposition is only valid up to trapping horizons, $f = 0$, where the Kodama and radial vectors are null, and the decomposition is no longer valid. The above requirement that at all times we are on an untrapped region, $f>0$, safeguards our analysis from that potential problem. Substituting the decomposition~\eqref{eq:k_decomposition} into the null equation \eqref{eq:null_geodesic} yields the decomposed null equation,
\begin{equation}
    - \Omega^2 + \Pi^2 + \frac{L^2}{r^2} = 0. \label{eq:null_geodesic_decomposition}
\end{equation}

The evolution of the time-radial components of the geodesic is determined by the evolution equation \eqref{eq:geodesic_evolution} contracted with a vector $X^\mu$, which after some manipulations provides the relation
\begin{equation}
    k^\nu \nabla_\nu \left( k_\mu X^\mu\right) = k^\mu k^\nu \nabla_\nu X_\nu.
\end{equation}
Then, the evolution equations of the energy and radial momentum along the geodesic are
\begin{align}
    \nabla_k \Omega & \equiv - \nabla_k \left( k_\mu u^\mu \right) = - k^a k^b u_{a:b}, \label{eq:evolution_energy} \\
    \nabla_k \Pi    &\equiv \nabla_k \left( k_\mu n^\mu \right) = k^a k^b n_{a:b} + \frac{L^2}{r^2} \sqrt{\frac{f}{r^2}}. \label{eq:evolution_radial}
\end{align}
These equations describe how the local normalized Kodama-frame energy and radial momentum evolve along the geodesic. The evolution of $\Omega$ depends on the contraction of the geodesic tangent with the covariant derivative of the normalized Kodama vector, $u_{a:b}$, while the evolution of $\Pi$ depends on the corresponding derivative of the radial vector, $n_{a:b}$, plus the term proportional to the angular momentum $L^2$, which we associate with the centrifugal force.

\section{Photon spheres in dynamical space-times} \label{sec:photon_surfaces}

In static space-times, the photon sphere is a well-defined static surface where null geodesics travel in unstable bound circular orbits. The final state of geodesics passing nearby the photon sphere is defined by comparing the impact parameter $b \equiv L/\omega$, where $\omega =-k_\mu\xi^\mu$ is the conserved energy associated with the time-like Killing vector $\xi^\mu$, with the critical impact parameter intrinsic to the space-time. For the static metric introduced in Eq.~\eqref{eq:static_metric}, this critical value is $b_c \equiv r_\text{ps}/\sqrt{h(r_\text{ps})}$. For orbits outside the photon sphere, if $b>b_c$ the orbits are scattered to asymptotic infinity, otherwise, $b<b_c$, the orbits fall to the inner region of the geometry (crossing the event horizon in the case of a black hole). The condition is the opposite from inside the photon sphere. In this sense, the photon sphere acts as the surface where the angular momentum flips its behavior. The main implication of these properties is the fact that the accessible space-time regions of a null geodesic with angular momentum $L\neq0$ are determined from constants of motion and, therefore, defined globally.

In a dynamical space-time, the above static photon sphere must be promoted to a quasi-locally defined surface, dubbed as the dynamical photon sphere. Due to the lack of a time-like Killing vector in these space-times the definition of such surfaces is troublesome. However, thanks to the definition of a Kodama vector and the decomposition in Eq.~\eqref{eq:k_decomposition}, we present below a covariant definition in which each instantaneous critical turning point $x_{\rm ps}\in\mathcal M_s$ selects an associated dynamical photon sphere $\mathcal S_{\rm ps}(x_{\rm ps})$ in the full space-time.

\subsection{Definition of dynamical photon spheres}

First, we define the \textit{turning points} in $\mathcal{M}_s$ as the set of points on the time-radial manifold, $x_m=\{x_m^a\}\in \mathcal{M}_s$, where the covariant radial momentum vanishes:
\begin{equation}
    \Pi(x_m) = 0. \label{eq:tp_mom}
\end{equation}
Each point $x_m \in \mathcal{M}_s$ forms, through spherical symmetry, a two-sphere in the full space-time. On these surfaces, the time-radial part of the wave vector is parallel to the normalized Kodama vector. Applying this condition to Eq.~\eqref{eq:null_geodesic_decomposition}, we find that the local normalized Kodama-frame energy at a turning point is related to the angular momentum and the areal radius by
\begin{equation}
    \Omega^2 \big|_{x_m} = \frac{L^2}{r^2} \bigg|_{x^a_m}.
\end{equation}
Rearranging this relation defines the local turning-point ratio $b_\text{loc}$ measured in the normalized Kodama frame:
\begin{equation}
    b_\text{loc}(x_m) \equiv \left|\frac{L}{\Omega}\right|_{x_m} = r(x_m). \label{eq:impact_variable}
\end{equation}
This local ratio is not a conserved impact parameter. In a general dynamical space-time there is no time-like Killing vector and, consequently, no conserved energy to define a global impact parameter. Even in a static space-time, the normalized Kodama-frame energy $\Omega$ is not conserved along a trajectory. The usual impact parameter must be defined using the conserved Killing energy, giving $b=L/\omega$.

In order to characterize the turning-point condition~\eqref{eq:tp_mom}, we use the decomposition in Eq.~\eqref{eq:k_decomposition} in the null geodesic equation~\eqref{eq:null_geodesic_decomposition}. The angular contribution proportional to $L^2$ is positive in the untrapped region, and Eq.~\eqref{eq:null_geodesic_decomposition} gives $\Omega^2\geq\Pi^2$. Radial turning is then defined by $\Pi=0$. For radial rays, $L=0$, there are no non-radial turning points of the type considered here. At a turning point, the time-radial projection $k_a$ is parallel to the Kodama direction, while the full wavevector retains its angular component.

We define a \textit{critical turning point} $x_\text{ps}\in\mathcal{M}_s$ as a point satisfying:
\begin{enumerate}[(i)]
    \item the turning-point condition~\eqref{eq:tp_mom}:
    \begin{equation}
        \Pi(x_\text{ps}) = 0.\label{eq:ps_mom}
    \end{equation}
    \item vanishing derivative of the radial momentum at that point:
    \begin{equation}
    \nabla_k \Pi \vert_{x_\text{ps}} = 0. \label{eq:ps_mom2}
    \end{equation}
\end{enumerate}

Each critical turning point $x_{\rm ps}$ selects the instantaneous dynamical photon sphere $\mathcal S_{\rm ps}(x_{\rm ps})\equiv\{x_{\rm ps}\}\times S^2$, obtained by spanning the angular coordinates over the symmetry two-sphere. As the space-time evolves, these spheres define a time-dependent family of instantaneous dynamical photon spheres. When this family varies smoothly, its union defines a photon sphere tube. This tube is not assumed to be a photon surface in the sense of Ref.~\cite{Claudel:2000yi} or to be generated by null geodesics.

Both conditions \eqref{eq:ps_mom} and \eqref{eq:ps_mom2} imply that the covariant radial momentum vanishes at those points as measured by the Kodama-defined time. Using the vector basis, we have that the turning points are defined by,
\begin{equation}
    k_a n^a = \Pi \vert_{x_m} = 0.
\end{equation}
At these points, the following expressions from Eqs.~\eqref{eq:evolution_energy} and~\eqref{eq:evolution_radial} are satisfied
\begin{align}
    \nabla_k \Omega &= 0 \label{eq:ps_energy}\\
    \nabla_k \Pi   &= - \frac{L^2}{r^2}  \left( \theta_n - \sqrt{\frac{f}{r^2}} \right). \label{eq:ps_mom3}
\end{align}
The first equation~\eqref{eq:ps_energy} implies that the instantaneous derivative of the normalized Kodama energy vanishes at a turning point, $\nabla_k \Omega = 0$, even in a dynamical space-time. Thus, the turning points of the null geodesic are stationary points of this local energy along the geodesic. As a remark, this does not imply that $\Omega$ is globally conserved.

The second equation~\eqref{eq:ps_mom3} describes the evolution of the radial momentum at the turning points. Then, using the definition of the critical turning points~\eqref{eq:ps_mom2}, we get
\begin{equation}
    \theta_n \vert_{x_\text{ps}} = \left[  \frac{\square r}{\sqrt{f}}- D_n \ln \sqrt{f} \right] _{x_\text{ps}} = \sqrt{\frac{f}{r^2}} \bigg\vert_{x_\text{ps}}. \label{eq:ps_mom4}
\end{equation}
This is the covariant equation of the location of the critical turning points, or equivalently the location of the dynamical photon spheres, using geometric quantities, equaling the expansion of the radial vector, $\theta_n$, with a particular value of $f/r^2$. As such, it is a generalization of the equation of the location of the photon sphere in a dynamical setting.  It depends exclusively on the radius $r$, and the metric function $f \equiv r_a r^a$, and their derivatives. 

Let us recap. Our covariant $2+2$ decomposition identifies a critical turning point $x_{\rm ps}\in\mathcal M_s$ through the instantaneous conditions $\Pi|_{x_{\rm ps}}=0$ and $\nabla_k\Pi|_{x_{\rm ps}}=0$, and associates it with the dynamical photon sphere $\mathcal S_{\rm ps}(x_{\rm ps})$. The definition is quasi-local, coordinate independent, and does not require a time-like Killing symmetry. As the space-time evolves, the associated two-spheres form a time-dependent family of instantaneous dynamical photon spheres and, when smooth, their union defines the photon sphere tube. In the static limit, the time-like Killing vector supplies a conserved energy and the location condition reduces to the standard static photon sphere.

Our definition has three main features:
\begin{enumerate}[(i)]
    \item  it uses coordinate-independent local covariant quantities,
    \item the definition does not depend on the existence of a time-like Killing symmetry,
    \item each critical turning point selects a dynamical photon sphere in the 4D space-time, allowing a time-dependent family in dynamical backgrounds.
\end{enumerate}
As a consequence of these definitions and features, in our analysis we are promoting the global static photon sphere to a quasi-local dynamical photon sphere definition.

One important point to note is that the existence of a photon sphere is independent of the existence of a trapping horizon. 
The normalized Kodama-radial frame is defined in the untrapped domain $f>0$, whereas a marginal sphere associated with a trapping horizon satisfies $f=0$. Consequently, the critical-turning-point equation may admit a photon sphere radius $r_{\rm ps}$ even when no trapping-horizon radius $r_{\rm h}$ exists. This allows for the formation of a photon sphere, for instance, within a collapsing star, before a trapping horizon appears. The existence of $\mathcal S_{\rm ps}(x_{\rm ps})$ alone does not establish finite-time trapping or the formation of observable photon rings, which would require null-geodesic evolution or ray tracing.

\subsection{Local radial deviation near the photon sphere}

In the previous section, we defined a dynamical photon sphere as the symmetry two-sphere associated with an instantaneous critical turning point of non-radial null motion. We now study the local radial deviation between neighboring orbits near that critical turning point and its associated dynamical photon sphere. This classification is local to a single event and does not by itself establish global stability along the time-dependent family.
 
To characterize this local radial deviation, we employ the geodesic deviation equation. Consider a reference null geodesic $\gamma(s)$ that passes through the critical turning point $x_\text{ps}$ with tangent vector $k_\mu$ and satisfies Eqs.~\eqref{eq:ps_mom} and~\eqref{eq:ps_mom2} at that event. We compare it with a neighboring null geodesic $\tilde{\gamma}(s)$ displaced by a small deviation vector $\xi_\mu$. At $x_\text{ps}$, this vector represents the separation between neighboring orbits near the instantaneous dynamical photon sphere. Its evolution is governed by the geodesic deviation equation~\cite{Misner:1973prb}:
\begin{equation}
    \frac{d^2 \xi^\mu}{ds^2} = - R^\mu_{\ \nu \sigma \rho } k^\nu \xi^\sigma k^\rho, \label{eq:jacobi}
\end{equation}
where $d/ds \equiv k^\mu \nabla_\mu$ is the covariant derivative along the ray, and $R^a_{\ bcd}$ is the Riemann curvature tensor of the space-time\footnote{We use the convention
\begin{equation}
    R^\rho_{\ \sigma\mu\nu} = \partial_\mu \Gamma^\rho_{\nu\sigma} - \partial_\nu \Gamma^\rho_{\mu\sigma} + \Gamma^\rho_{\mu\lambda}\Gamma^\lambda_{\nu\sigma} - \Gamma^\rho_{\nu\lambda}\Gamma^\lambda_{\mu\sigma}.
\end{equation}.}.

At the single event $x_\text{ps}$, the reference geodesic satisfies the critical-turning-point conditions~\eqref{eq:ps_mom} and~\eqref{eq:ps_mom2}. Consequently, its time-radial momentum $k_a$ is proportional to the normalized Kodama vector $u_a$ at that point:
\begin{equation}
    k_\mu \vert_{x_\text{ps}} = \left( \Omega u_a, k_A \right)\vert_{x_\text{ps}}.
\end{equation}
Then, we consider a deviation vector that is orthogonal to the orbit and aligned with the radial direction: 
\begin{equation}
    \xi_\mu = \left( \xi n_a, 0 \right),
\end{equation}
where $\xi(s)$ is the scalar magnitude of the deviation. After lengthy but straightforward algebra, we find that the radial projection of the deviation equation~\eqref{eq:jacobi}, evaluated at the single critical-turning event $x_\text{ps}$, reduces to:
\begin{equation}
    \nabla_k \nabla_k \xi - \mathcal{K} \xi = 0, \label{eq:xi_ode}
\end{equation}
where 
\begin{equation}
    \mathcal{K} \equiv - \frac{1}{f}\left( R_{\mu \nu \alpha \beta}r^\mu k^\nu r^\alpha k^\beta\right).
\end{equation}
The quantity $\mathcal K$ is the radial tidal curvature experienced by a geodesic near the photon sphere in the $r_a$ direction.
In the normalized basis, this curvature takes the geometrical form
\begin{align}
    \mathcal{K}&= \frac{L^2}{r^2} \left(  \frac{\mathcal{R}}{2} + \frac{\sqrt{f}}{r}  D_n \ln \sqrt{f} \right)\bigg \vert_{x_\text{ps}}. \label{eq:tidal_curvature}
\end{align}
where $\mathcal{R}$ is the 2-dimensional curvature scalar of the time-radial manifold.

An analytical solution to the deviation equation \eqref{eq:xi_ode} is possible only when $\mathcal{K}$ is constant along the geodesic, something which occurs in static space-times. In a general dynamical case, the sign of $\mathcal{K}$ at a critical turning point characterizes only the local radial deviation between neighboring null geodesics at that point: \begin{itemize}
    \item If $\mathcal{K} > 0$, the local radial deviation is separating.  If $\mathcal{K}$ is positive and constant over a finite interval, Eq.~\eqref{eq:xi_ode} admits exponentially growing and decaying solutions, with $\Lambda=\sqrt{\mathcal{K}}$. The quantity $\Lambda$ is usually referred to as the {\it Lyapunov exponent}.
    \item If $\mathcal{K} < 0$, the local radial deviation is restoring. If $\mathcal{K}$ remains negative and constant over a finite interval, Eq.~\eqref{eq:xi_ode} becomes a harmonic-oscillator equation with frequency $\omega=\sqrt{-\mathcal{K}}$.
    \item If $\mathcal{K} = 0$, the local radial deviation is marginal. If $\mathcal K$ remains zero through an interval, perturbations evolve linearly, $\xi \sim as + b$. \end{itemize}

In a non-stationary space-time, determining whether neighboring geodesics remain bounded or separate over a finite interval requires evolving the full deviation equation along a reference geodesic. The sign of $\mathcal{K}$ at one critical turning point is not a complete proof of global orbital stability.

\subsection{Effect of the source on the photon sphere}

Until now, we have derived the location of the photon sphere and the local radial deviation between neighboring geodesics using only geometrical quantities that depend on the two-dimensional metric tensor $g_{ab}$ and the areal radius $r(x^a)$. However, in physical scenarios, the geometry of the space-time is determined by a matter source. In this section, we derive the same equations for the photon sphere considering that the space-time is threaded by a general source tensor.

We use geometrized units $G=1$ throughout. The field equations are written as
\begin{equation}
 G_{\mu\nu} = S_{\mu \nu},
\end{equation}
where $G_{\mu\nu}$ is the Einstein tensor and $S_{\mu\nu}=8\pi T_{\mu\nu}$ is the geometrically normalized source tensor. A source that respects the spherical symmetry of the space-time can be decomposed as
\begin{equation}
    S_{\mu \nu} = 
    \begin{bmatrix}
        S_{ab} & 0 \\
        0 & P_\perp r^2 \Omega_{AB}
    \end{bmatrix},
\end{equation}
where the time-radial part is expanded on the basis $\{u_a, n_a\}$ as
\begin{equation}
    S_{ab} = E u_a u_b + P n_a n_b + 2 Q u_{(a} n_{b)} .
\end{equation}
The geometric scalars $E$, $P$, $Q$, and $P_\perp$ are functions on $\mathcal{M}_s$ that characterize the source.
To find their physical meaning, we project the source tensor onto an observer's reference frame defined by the orthonormal frame vectors $\{u_a, n_a\}$. Then, the physical energy density $\varepsilon$, radial pressure $p_r$, and radial energy flux $q$, as measured by this observer, are given by:

\begin{align}
    8 \pi \varepsilon &\equiv S_{ab} u^a u^b  = E, \\
    8 \pi p_r       &\equiv S_{ab} n^a n^b  = P, \\
    8 \pi q         &\equiv -S_{ab} u^a n^b = Q.
\end{align}
Thus, $E$, $P$, and $Q$ are respectively $8\pi$ times the physical energy density, radial pressure, and radial energy flux measured in the normalized Kodama frame. Likewise, $P_\perp=8\pi p_t$. From these, one can solve the Einstein field equations for this general source. The complete derivation is detailed in Appendix~\ref{sec:gravitational_field_equations}.

Using the Einstein field equation~\eqref{eq:r_ab}, we express the location of the photon sphere dependent on the matter content as
\begin{equation}
        \left[ r- 3m - \frac{1}{2} r^3 P\right]_{x_\text{ps}} = 0. \label{eq:ps_condition}
\end{equation}
This relation only depends on the areal radius $r$, the Misner-Sharp mass $m$, and the local radial pressure source $P$. It generalizes its counterpart of the static limit, which was  derived in \cite{Claudel:2000yi}, to provide an equivalent expression that is valid in a dynamical setup where all these quantities evolve in time.

In global dynamical photon surface approaches~\cite{Claudel:2000yi}, the photon surface location $r_\text{ps}$ is selected by integrating a null geodesic with a late-time or stationary boundary condition. Our algebraic equation~\eqref{eq:ps_condition} uses local geometric and matter data to describe the position of instantaneous critical turning points. We expect the two approaches to yield similar behavior in an adiabatic regime, where the space-time evolves slowly compared with the orbital timescale. In the Vaidya example of Sec.~\ref{sec:examples}, we make the corresponding slow-transfer limit $|\dot m|\ll1$ explicit and interpret our result within this quasi-local approximation.

In addition, we can use the matter content to express the physical conditions for the local radial deviation near the photon sphere. Using the Einstein equations and the photon sphere equation \eqref{eq:ps_condition} we get
\begin{equation} \label{eq:extmatt}
    \mathcal{K} \vert_{x_\text{ps}} = \left[ \frac{L^2}{r^4} \left( 1  - \left(E + P_\perp\right)r^2 \right) \right]_{x_\text{ps}}.
\end{equation}
From this relation, we can determine the local radial deviation near the photon sphere for a given source. The local radial deviation between neighboring geodesics is restoring if $\mathcal{K}<0$, separating if $\mathcal{K}>0$, and marginal if $\mathcal{K}=0$. Therefore,
\begin{align*}
    \left[(E + P_\perp) r^2 \right]_{x_\text{ps}} &> 1 & \text{locally restoring}\\
    \left[(E + P_\perp) r^2 \right]_{x_\text{ps}} &< 1 & \text{locally separating}
\end{align*}

For completeness of our analysis, let us assume now that the matter source is a physical energy-momentum tensor satisfying the Weak Energy Condition (WEC). This condition implies that the local energy density plus the tangential pressure must be non-negative,
\begin{equation}
    E + P_\perp \geq 0.
\end{equation}
As a consequence, under the WEC, a physical source producing a locally separating radial deviation at the critical turning point must satisfy the inequalities
\begin{equation}\label{eq:matter_stability_condition}
    0 \leq \left[(E + P_\perp) r^2 \right]_{x_\text{ps}} < 1.
\end{equation}
For static space-times, the distinction between stable and unstable photon spheres is critical to determine the stability of the space-time itself. It has been argued in the literature, see e.g. ~\cite{Cardoso:2014sna,Keir:2014oka,Cunha:2017qtt,Cunha:2022gde}, that space-times with stable photon spheres (such as a sufficiently-compact horizonless object \cite{Cardoso:2019rvt}) are prone to exhibit long-lived trapping of perturbations, which can lead to non-linear instabilities triggering the destabilization of the corresponding object. This would seemingly demand the photon spheres of a given space-time to be unstable, restricting the physical matter to the bounds expressed in Eq.~\eqref{eq:matter_stability_condition}. However, in recent times the onset and development of such instabilities has been refined to show that they depend sensitively on the underlying space-time geometry and field dynamics \cite{Zhong:2022jke,Guo:2024cts}, implying that such objects are not automatically ruled out from the point of view of astronomical considerations.
In a dynamical space-time, the local bound~\eqref{eq:matter_stability_condition} characterizes only a locally separating radial deviation at a critical turning point and cannot, without an evolution analysis, be used to infer either long-lived trapping or stability of the background.

We emphasize that the classification above is local and instantaneous: the sign of $\mathcal K$ at the critical turning point $x_{\rm ps}$ characterizes the local radial deviation between neighboring null geodesics at the associated dynamical photon sphere $\mathcal S_{\rm ps}(x_{\rm ps})$. This is different from global stability, which depends on the evolution over a finite interval of affine parameter and, in a dynamical space-time, on the variation of $\mathcal K$ along the geodesic. Establishing global stability, long-lived trapping, or an instability of the background space-time therefore requires solving the corresponding evolution problem beyond the local condition derived here.

\section{Examples}\label{sec:examples}

In this section, we present the derivation and discuss the properties of the quasi-local dynamical photon spheres on different space-times. We first shall calculate the stationary limit in a Schwarzschild-like solution to make contact with well known expressions. Then, we apply this to dynamical cases of interest, such as stellar collapse or radiating black holes.

\subsection{Static Schwarzschild-like solutions}

We consider a general static Schwarzschild-like spacetime, with the areal radius $r$ as the standard radial coordinate
\begin{equation}\label{eq:static_metric}
    \der s^2 = - h(r) \der t^2 + 1/f(r) \der r^2 + r^2 \der \Omega^2,    
\end{equation}
where we have already defined the metric function $f(r)$ in Eq. (\ref{eq:def_f}), 
and we keep $h(r) \neq f(r)$ for general matter contents and not exclusively black hole solutions.

For such coordinates, the normalized Kodama and radial vectors are defined in the time-radial manifold as
\begin{align*}
    u_a &\equiv \left( -\sqrt{h(r)}, 0\right), & n_a &\equiv \left(0, 1/\sqrt{f(r)} \right), \\
    u^a &= \left( 1/ \sqrt{h(r)}, 0 \right), & n^a &= \left(0, \sqrt{f(r)} \right).
\end{align*}
Then, the derivatives on a scalar field $\phi(r)$ are
\begin{align*}
    D_u \phi &\equiv u^a \nabla_a \phi(r) =  \frac{1}{\sqrt{h(r)}} \partial_t \phi, \\
    D_n \phi &\equiv n^a \nabla_a \phi(r) = \sqrt{f(r)} \partial_r \phi,
\end{align*}
where the prime indicates the partial derivative with respect to the radial coordinate. On the other hand, from the metric (\ref{eq:static_metric}), we calculate the main kinematic quantities:
\begin{align}
    \theta_u &= 0, \\
    \theta_n &= \sqrt{f} \frac{h'(r)}{2h(r)}.
\end{align}
Therefore, by using Eq. \eqref{eq:ps_mom4}, we find the location of the photon sphere in this case as
\begin{equation} \label{eq:ps}
    \left[ \frac{h'(r)}{h(r)}  - \frac{2}{r} \right]_{r_\text{ps}}= 0 \ ,
\end{equation}
which coincides with the standard relation for the location of the static photon sphere in this Schwarzschild-like space-time. For instance, for the usual Schwarzschild black hole, $h(r)=f(r)=1-2M/r$, this yields the well known result $r_{\text{ps}}=3M$.

The instability of orbits around the photon sphere is determined by the tidal curvature $\mathcal{K}$ as defined in Eq. \eqref{eq:tidal_curvature}. For the static metric (\ref{eq:static_metric}) this reads as

\begin{equation}
   \mathcal{K}_{\text{ps}} \equiv  \mathcal{K}\vert_{r_\text{ps}} = \left[\frac{L^2}{r^2} \frac{f}{2h} \left( \frac{2h}{r^2} - h''\right)  \right] \bigg \vert_{r_{\text{ps}}}
\end{equation}

For $\mathcal{K}>0$, it coincides with the Lyapunov exponent $\Lambda_\text{ps}$ for a static space-time via the normalization
\begin{equation}
    \Lambda_\text{ps} \equiv \sqrt{ \left| \mathcal{K}_{\text{ps}} \right| },
\end{equation}
while for $\mathcal{K}<0$, $\sqrt{-\mathcal{K}}$ is the stable oscillation frequency.
This exponent represents a measure of the instability scale (in affine time) of nearly-bound geodesics in this static case. Typically, in phenomenological settings such as in the imaging of supermassive objects or in quasi-normal modes in the ringdown phase of binary mergers, it is more convenient to express this exponent in terms of the equation of either the affine time $t_{\text{ps}}$ or the angular deflection $\phi_{\text{ps}}$ experienced by such orbits. This can be easily done by appealing to the conservation laws holding in such a static case. In this sense, using the energy conservation equation and after some manipulations involving the critical impact parameter, which in this case reads as $b_{\text{ps}} =\tfrac{r_{\text{ps}}}{\sqrt{h_{\text{ps}}}}$, and the photon sphere condition (\ref{eq:ps}), one can write
\begin{equation}
\lambda_{\text{ps}}=\frac{h}{\omega} \bigg \vert_{r_{\text{ps}}} \Lambda_{\text{ps}} = \frac{1}{\sqrt{2}} \sqrt{\frac{f_{\text{ps}}}{r_{\text{ps}}^2} (2h_{\text{ps}}-r_{\text{ps}}^2 h_{\text{ps}}'')}
\end{equation}
Note that this expression agrees with Cardoso's result (using our notation), originally found in \cite{Cardoso:2008bp}. This exponent encodes the time-scale of the perturbations around the photon sphere via $t_{\text{ps}} \propto \lambda_{\text{ps}}^{-1}$, which may be of phenomenological interest in transient observations (hot-spots) in magnetized accretion environments \cite{Ripperda:2020bpz,Vos:2022yij}. For completeness, let us mention that using the conservation of the angular momentum we can transform this exponent into
\begin{equation}
\gamma_{\text{ps}}=\pi b_{\text{ps}} \lambda_{\text{ps}}
\end{equation}
This new version of the Lyapunov exponent defines the instability scale in the azimuthal angle or, equivalently, in the number of {\it half-turns} performed by photon trajectories around a compact object. As such, it provides a direct link between highly-lensed trajectories in the black hole gravitational field and the creation of photon rings in the observer's plane image in time-averaged images \cite{Gralla:2019xty,Johnson:2019ljv,Vincent:2022fwj,Staelens:2023jgr}.

Alternatively, we may determine the location of the photon sphere using the matter content of the solution. To this end, let us consider a static source (no off-diagonal flux term in the source tensor, $Q=0$), as given by

\begin{equation}
    S_{ab} = 8\pi\left(\varepsilon u_a u_b + p_r n_a n_b\right), \quad S_{AB} = 8\pi p_t r^2 \Omega_{AB},
\end{equation}

where we have used a coordinate observer representing the fluid velocity $u^\mu$. The Misner-Sharp mass appearing in Eq. (\ref{eq:def_f}) is ultimately determined by the energy source via the equations
\begin{align}
    \partial_r m(r) &= 4\pi r^2 \varepsilon(r), \\
    \partial_t m(r) &= 0.
\end{align}
Thus,

\begin{equation} \label{eq:mss}
 m(r) = M + 4\pi \int^r s^2 \varepsilon(s) \der s.
\end{equation}
where $M$ is an integration constant.  The location of the trapping surfaces in this setup, found by solving $f(r_{\text{h}}) = 1 - 2m(r_{\text{h}})/r_{\text{h}} = 0$, are defined by

\begin{equation} \label{eq:evma}
    2 M + 8 \pi \int^{r_{\text{h}}} s^2 \varepsilon(s) \der s - r_{\text{h}} = 0.
\end{equation}

while the location of the photon sphere, as given by Eq. (\ref{eq:ps_condition}),  is given by

\begin{equation} \label{phfluid}
    3 M + 12\pi \int^{r_{\text{ps}}} s^2 \varepsilon(s) \der s + 4\pi r_{\text{ps}}^3 p_r(r_{\text{ps}}) - r_{\text{ps}} = 0.
\end{equation}

To illustrate this procedure, let us consider the Reissner-Nordstr\"om solution that describes a spherically symmetric black hole characterized by mass $M$ and electric charge $Q_e$. The electromagnetic source has physical energy density and pressures
\begin{equation*}
    \varepsilon(r) = \frac{Q_e^2}{8\pi r^4} = - p_r(r) = p_t(r).
\end{equation*}
which recovers the Schwarzschild solution in the $Q_e\to 0$ limit. Using the above expressions, the Misner-Sharp mass (\ref{eq:mss}) in this case is given by:
\begin{equation}
    m(r) = M - \frac{Q_e^2}{2r}.
\end{equation}
The trapping horizons, given by Eq. (\ref{eq:evma}),  read as
\begin{equation}
    r_\pm = M \pm \sqrt{M^2 - Q_e^2}.
\end{equation}
with the larger one corresponding to the event horizon, namely, $r_+=r_{\text{h}}$. The location of the photon sphere, $r_{\text{ps}}$, as determined by Eq. (\ref{phfluid}), reads as 
\begin{equation}
    r_\text{\text{ps}}^2 - 3M r_\text{\text{ps}} + 2Q_e^2 = 0,
\end{equation}
whose solutions are
\begin{equation}
    r_\text{ps} = \frac{3M \pm \sqrt{9M^2 - 8Q_e^2}}{2}. \label{eq:RN_ps}
\end{equation}
These solutions give the radius of the inner and outer photon spheres and recover the Schwarzschild value as $Q_e\to0$. However, the inner root lies inside the outer horizon and therefore falls outside the domain considered here. 

Finally, we calculate the tidal curvature $\mathcal{K}$ from the matter equation \eqref{eq:extmatt}, which yields
\begin{equation}
    \mathcal{K}_{\text{ps}} = \left[\frac{L^2}{r^4} \left(1  - \frac{2Q_e^2}{r^2} \right) \right] \bigg \vert_{r_\text{ps}}.
\end{equation}
For Reissner--Nordstr\"om black holes, $M\geq|Q_e|$, the tidal curvature of the outer photon sphere is positive, so the orbit is unstable. The weaker condition $9M^2\geq8Q_e^2$ guarantees real photon-sphere roots accordingly to Eq.~\eqref{eq:RN_ps}. 

\subsection{Stellar collapse}

We next study the formation of a dynamical photon sphere  and the behaviour of nearly-bound orbits of a star collapsing onto a black hole. We use the Oppenheimer-Snyder (OS) model which describes the gravitational collapse of a pressureless and homogeneous, spherically symmetric fluid.

The space-time is divided into two regions: an exterior vacuum Schwarzschild region and an interior region described by a Friedmann-Lemaitre-Robertson-Walker (FLRW) metric. Both space-times are matched at the surface of the fluid, where the corresponding areal radius, $R$, and the Misner-Sharp mass, $m$, are continuous along each space-time. We use these constraints to match both solutions at the surface.

The exterior space-time is the static Schwarzschild solution in $(t, r, \theta, \phi)$ coordinates, that is:
\begin{equation}
    \der s^2_\text{e} = - \left( 1 - \frac{2M}{r}\right) \der t^2 + \frac{1}{ \left( 1 - \frac{2M}{r}\right)} \der r^2 + r^2 \der \Omega^2,
\end{equation}
where the areal radius and the Misner-Sharp mass are given by
\begin{equation}
    R_\text{e}(r) = r, \quad m_\text{e} = M.
\end{equation}
The interior metric, in comoving coordinates~$(\tau, \chi, \theta, \phi)$, is
\begin{equation}
    \der s^2_\text{i} = -\der \tau^2 + a^2(\tau) \left[ \der\chi^2 + \sin^2\chi \der\Omega^2 \right],
\end{equation}
where $a(\tau)$ is the scale factor. For this metric, the areal radius and the Misner-Sharp mass are given by
\begin{align}
    R_\text{i}(\tau, \chi) &= a(\tau) \sin\chi, \\
    m_\text{i}(\tau, \chi) &= \frac{4\pi}{3} R_\text{i}^3(\tau, \chi) \varepsilon(\tau),
\end{align}
respectively, where we have integrated the mass and used the Friedmann equation,
\begin{equation}
    1 + (\partial_\tau a(\tau))^2 = \frac{8\pi}{3} a^2(\tau) \varepsilon(\tau).
\end{equation}

At the fluid boundary we have the matching surface between space-times, denoted by $b$, for which the following conditions hold
\begin{align}
    R_\text{i}\vert_b &= R_\text{e}\vert_b \implies r_b(\tau) = a(\tau)\sin\chi_b, \\
    m_\text{i}\vert_b &= m_\text{e}\vert_b \implies M = \frac{4\pi}{3} R_\text{i}^3(\tau, \chi_b) \varepsilon(\tau) .
\end{align}
From these two relations and the definitions, the interior mass and radius can be related to the boundary quantities as
\begin{equation}\label{eq:os_mass_radius}
    \frac{m_i}{R_i^3} = \frac{M}{R_b^3}.
\end{equation}

\subsubsection{Trapping horizons and photon spheres}

Now, we calculate the characteristic surfaces of the space-time. We start by characterizing the trapping horizons defined by the trapping surface equation~\eqref{eq:trap_surface}.

On the vacuum region outside the star, the space-time is described by the static Schwarzschild metric. Therefore, the trapping horizon is constant and coincides with the event horizon\begin{equation}
    r_\text{EH} = 2M.
\end{equation}
This is only valid when the surface of the star has contracted below this point, $r_b < 2M$. In this external region, where $r > a(\tau)\sin\chi_b$, the mass is constant, $m=M$. The condition becomes $r - 2M = 0$. Thus, outside the star, the apparent horizon coincides with the event horizon at $r=2M$.

Inside the star, where $r < a(\tau)\sin\chi_b$, the Misner-Sharp mass is a function of the radius, $m(r, \tau) = M (r / [a(\tau)\sin\chi_b])^3$. Using Eq.~\eqref{eq:os_mass_radius}, the condition for the apparent horizon becomes:
\begin{equation}
    r = 2m(r, \tau) \implies r = 2M \left( \frac{r}{a(\tau)\sin\chi_b} \right)^3.
\end{equation}
Solving for $r$ gives the radius of the interior apparent horizon, $r_\text{AH}(\tau)$:
\begin{equation}
    r_\text{AH}(\tau)^2 = \frac{\left(a(\tau)\sin\chi_b\right)^3}{2M}.
\end{equation}
The apparent horizon first forms at the boundary of the star when the horizon crosses it, at $r_\text{EH}$, then it contracts to the central singularity as the collapse continues.

Once we have the trapping horizons located, we next calculate the photon spheres of the collapsing star. For a pressureless dust ($p=0$), our general condition for the photon sphere, Eq.~\eqref{eq:ps_condition}, simplifies to $r - 3m(r, \tau) = 0$. In the vacuum exterior, this simply gives the static photon sphere at $r_{\text{ps}} = 3M$. 

Inside the star, a dynamical photon sphere can exist. The condition $r=3m(r,\tau)$ selects a time-dependent family of instantaneous dynamical photon spheres with radius
\begin{equation}
    r_\text{ps}^2(\tau) = \frac{(a(\tau)\sin\chi_b)^3}{3M},
\end{equation}
or using the total mass at the boundary
\begin{equation}
    r_\text{ps}^2(\tau) = \frac{1}{4 \pi \varepsilon(\tau) }. \label{eq:os_model_raps}
\end{equation}
This family first appears at the boundary when the stellar surface crosses the static photon sphere $r=3M$; its radii then decrease toward the singularity, tracing the interior photon sphere tube.

The local radial deviation near the photon sphere inside the star is characterized by
\begin{equation}
    \mathcal{K}_{ps} = \left[ \frac{L^2}{r^4} \left( 1  - 8 \pi \varepsilon(\tau) r^2 \right) \right] \bigg \vert_{r_\text{ps}}.
\end{equation}
Using the specific relations for the OS model \eqref{eq:os_model_raps}, we get
\begin{equation}
    \mathcal{K}_{ps} = -  4 \pi \varepsilon(\tau) \frac{L^2}{r_\text{ps}^2},
\end{equation}
which is always negative provided that the energy density is positive, $\varepsilon(\tau) > 0$. 
Thus, at each critical turning point, the local radial deviation between neighboring null geodesics near the instantaneous dynamical photon sphere is restoring. This local sign does not establish bounded orbital evolution along the photon sphere tube. As the photon sphere radius approaches the singularity, $r_\text{ps}\to0$, the absolute value of the tidal curvature diverges.

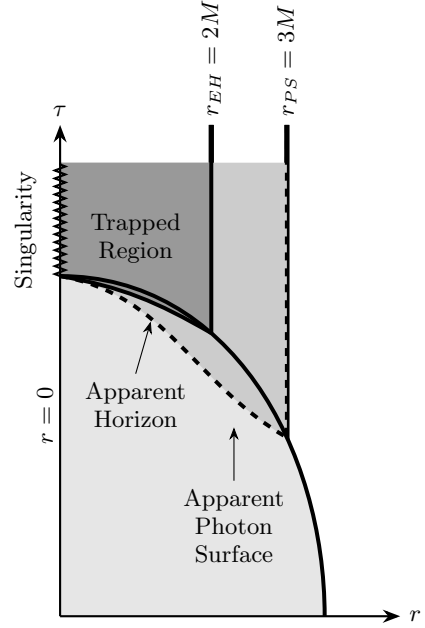
\begin{figure}[t!]
    \begin{tikzpicture}[scale=1, >=Stealth, font=\small]

        \def\Rinit{3.5}          \def\Rhorizon{2}       \def\Rphoton{3}       \def\Tcollapse{4.5}     \def\Hmax{6.5}         \def\Hoffset{0.5}

        \coordinate (O) at (0,0);
    \coordinate (R_axis_end) at (4.5,0);
    \coordinate (T_axis_end) at (0,\Hmax);
    \coordinate (StarSurfaceInit) at (\Rinit, 0);
    \coordinate (SingularityStart) at (0, \Tcollapse);
    \coordinate (HorizonTop) at (\Rhorizon, \Hmax);
    \coordinate (SingularityTop) at (0, \Hmax-\Hoffset);
    \coordinate (PhotonTop) at (\Rphoton, \Hmax);

        \draw[name path=event_horizon,line width=2pt] (HorizonTop) -- (\Rhorizon, 0);
    \draw[name path=photon_surface, line width=2pt] (PhotonTop) -- (\Rphoton, 0);

    \draw [name path=boundary, thick] (SingularityStart)
        to [out=0, in= 90] (StarSurfaceInit);

    \path[name intersections={of=event_horizon and boundary, by=eh_bound}];
    \path[name intersections={of=photon_surface and boundary, by=ps_bound}];

        \fill [black!10] (SingularityStart)
        to [out=0, in= 90] (StarSurfaceInit)
        to (O);

    \fill [black!20] (O)
        to [] (ps_bound)
        to [out=90, in= -90] (\Rphoton, \Hmax-\Hoffset)
        to [out=180, in= 0]  (0, \Hmax-\Hoffset)
        to (SingularityStart)
        to[out=-10, in=150] (ps_bound);

    \fill [black!40] (O)
        to [] (eh_bound)
        to [out=90, in= -90] (\Rhorizon, \Hmax-\Hoffset)
        to [out=180, in= 0]  (0, \Hmax-\Hoffset)
        to [] (SingularityStart)
        to[out=-10, in=150] (eh_bound);

        \draw[line width=1.5pt] (HorizonTop) -- (eh_bound);
    \draw[dashed,line width=1.5pt] (\Rphoton, \Hmax) -- (ps_bound);

    \draw[line width=1.5pt] (SingularityStart) to[out=-10, in=150] (eh_bound);
    \draw[dashed, line width=1.5pt] (SingularityStart) to[out=-10, in=150] (ps_bound);

    \draw [line width=1.5pt] (SingularityStart)
        to [out=0, in= 90] (StarSurfaceInit);

        \draw[->, thick] (O) -- (R_axis_end) node[right] {$r$};
    \draw[->, thick] (O) -- (T_axis_end) node[above] {$\tau$};

        \draw[decorate, decoration={zigzag, amplitude=2pt, segment length=3pt}, thick] 
        (SingularityStart) -- (SingularityTop);
        \node[above, align=center, anchor= south east, rotate=90, anchor=west] at (-0.2, \Hmax/3) {$r=0$};

    \node[below, rotate=90, anchor=west] at (HorizonTop) {$r_{EH}=2M$};
    \node[below, rotate=90, anchor=west] at (PhotonTop) {$r_{PS}=3M$};

        \node[align=center, anchor=south east, rotate=90] (SingularityLabel) at (-0.2, \Hmax-\Hoffset) {Singularity};

        \node[align=center, anchor=center] (HorizonLabel) at (1, \Hmax - 3*\Hoffset) {Trapped\\Region};

        \node[align=center, anchor=center] (ApparentLabel) at (1.0, 2.8) {Apparent\\Horizon};
        \draw[->] (ApparentLabel.north) -- (1.2, 3.9); 

        \node[align=center, anchor=center] (ApparentPSLabel) at (2.3, 1.2) {Dynamical photon sphere};
        \draw[->] (ApparentPSLabel.north) -- (2.3, 2.5); 
                    \end{tikzpicture}
    \caption{Diagram of the Oppenheimer-Snyder collapse, including the event horizon and the photon sphere. We graph with a solid line the trapping horizon, and the photon sphere tube with a dashed line.}
        \label{fig:OS}
\end{figure}

\subsubsection{Formation of the horizon and the photon sphere tube}

The OS model provides an idealized picture of how a time-dependent family of dynamical photon spheres defines a photon sphere tube in a collapsing space-time. We graph a diagram of the Oppenheimer-Snyder collapse in Fig.~\ref{fig:OS}.

Initially, when the star's radius is greater than $3M$, the exterior space-time contains no photon sphere, and no light-trapping surfaces exist anywhere. No null orbit is trapped in circular motion. Eventually, the boundary collapses beyond $r<3M$, crossing the photon sphere. Now, the space-time outside the fluid features a critical impact parameter, $b_{    \text{ps}} = 3 \sqrt{3}M$, that determines the escaping condition for the null orbits. Orbits from outside the photon sphere with $b< b_{\text{ps}}$ will be locally trapped, the opposite condition, $b>b_{\text{ps}}$, affects orbits from inside the photon sphere.
Later in the collapse, when the boundary contracts past $r_b = 2M$, an apparent horizon forms at the surface and falls inward towards the singularity. Once both tubes lie inside the fluid, their instantaneous radii satisfy $r_{\rm ps}<r_{\rm AH}$. Since the trapped region in the homogeneous OS interior is the outer region $r_b>r>r_{\rm AH}$, each instantaneous dynamical photon sphere lies in the untrapped region $r<r_{\rm AH}$, where our derivation remains valid.

The OS model has a drawback. The formation of the apparent horizon starts at the surface and not at the center, where the higher density in the center should trigger the formation, to expand until matching the event horizon on the exterior. The same issue occurs for the first appearance of the time-dependent family of dynamical photon spheres. This is a well known subtlety of the OS model \cite{Joshi:1987wg}. It is related to the fact that the OS model has a homogeneous matter density. As the radial gradient of matter is zero, there is no way to transfer causal information from the center to the surface. 

\subsubsection{Lemaitre-Tolman-Bondi collapse}

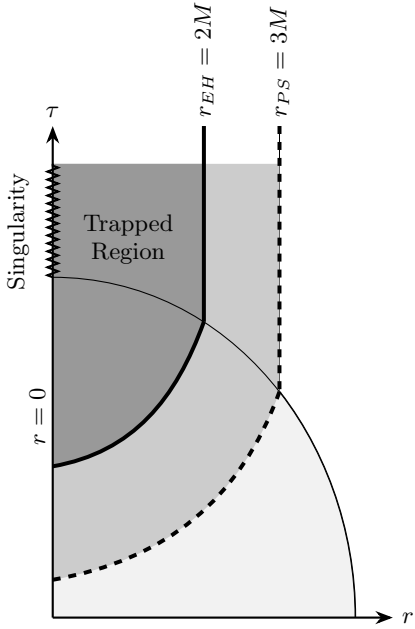
\begin{figure}[t!]
    \begin{tikzpicture}[scale=1, >=Stealth, font=\small]

        \def\Rinit{4}          \def\Rhorizon{2}       \def\Rphoton{3}       \def\Tcollapse{4.5}     \def\Hmax{6.5}         \def\Hoffset{0.5}
    \def\Thorizon{2}
    \def\Tphoton{0.5}

        \coordinate (O) at (0,0);
    \coordinate (R_axis_end) at (4.5,0);
    \coordinate (T_axis_end) at (0,\Hmax);
    \coordinate (StarSurfaceInit) at (\Rinit, 0);
    \coordinate (SingularityStart) at (0, \Tcollapse);
    \coordinate (HorizonTop) at (\Rhorizon, \Hmax);
    \coordinate (SingularityTop) at (0, \Hmax-\Hoffset);
    \coordinate (PhotonTop) at (\Rphoton, \Hmax);
    \coordinate(HorizonStart) at (0, \Thorizon);
    \coordinate(PhotonStart) at (0, \Tphoton);

        \draw[name path=event_horizon,line width=0pt] (HorizonTop) -- (\Rhorizon, 0);
    \draw[name path=photon_surface, line width=0pt] (PhotonTop) -- (\Rphoton, 0);

    \draw [name path=boundary, thick] (SingularityStart)
        to [out=0, in= 90] (StarSurfaceInit);

    \path[name intersections={of=event_horizon and boundary, by=eh_bound}];
    \path[name intersections={of=photon_surface and boundary, by=ps_bound}];

        \fill [black!5] (SingularityStart)
        to [out=0, in= 90] (StarSurfaceInit)
        to (O);

    \fill [black!20] (O)
        to [] (ps_bound)
        to [out=90, in= -90] (\Rphoton, \Hmax-\Hoffset)
        to [out=180, in= 0]  (0, \Hmax-\Hoffset)
        to (PhotonStart)
        to[out=10, in=250] (ps_bound);

    \fill [black!40] (O)
        to [] (eh_bound)
        to [out=90, in= -90] (\Rhorizon, \Hmax-\Hoffset)
        to [out=180, in= 0]  (0, \Hmax-\Hoffset)
        to [] (HorizonStart)
        to[out=10, in=250] (eh_bound);

        \draw[line width=1.5pt] (HorizonTop) -- (eh_bound);
    \draw[dashed, line width=1.5pt] (\Rphoton, \Hmax) -- (ps_bound);
    
            \draw[line width=1.5pt] (HorizonStart) to[out=10, in=250] (eh_bound);
    \draw[dashed, line width=1.5pt] (PhotonStart) to[out=10, in=250] (ps_bound);

    \draw [] (SingularityStart)
        to [out=0, in= 90] (StarSurfaceInit);

        \draw[->, thick] (O) -- (R_axis_end) node[right] {$r$};
    \draw[->, thick] (O) -- (T_axis_end) node[above] {$\tau$};

        \draw[decorate, decoration={zigzag, amplitude=2pt, segment length=3pt}, thick] 
        (HorizonStart) -- (SingularityTop);
        \node[above, align=center, anchor= south east, rotate=90, anchor=west] at (-0.2, \Hmax/8) {$r=0$};

        \node[below, rotate=90, anchor=west] at (HorizonTop) {$r_{EH}=2M$};
    \node[below, rotate=90, anchor=west] at (PhotonTop) {$r_{PS}=3M$};

        \node[align=center, anchor=south east, rotate=90] (SingularityLabel) at (-0.2, \Hmax-\Hoffset) {Singularity};

        \node[align=center, anchor=center] (HorizonLabel) at (1, \Hmax - 3*\Hoffset) {Trapped\\Region};
\end{tikzpicture}

    \caption{Diagram of the Lemaitre-Tolman-Bondi collapse, including the event horizon and the photon sphere tube.}
    \label{fig:ltb}
\end{figure}

A better physical model is the Lemaitre-Tolman-Bondi (LTB) collapse. It assumes pressureless fluid, but an inhomogeneous matter density. The model solves the causality problems associated to the formation of the surfaces. The apparent trapping horizon appears in the center and expands to the surface at $R=2M$. Equivalently, the time-dependent family of instantaneous dynamical photon spheres begins at the center and reaches the static photon sphere radius $r=3M$ at the stellar surface. The corresponding photon sphere tube depends on the Misner--Sharp mass and therefore requires a specified energy-density profile. Because the LTB model is pressureless, the algebraic radii $r_{\rm ps}=3m$ and $r_{\rm AH}=2m$, evaluated for the same local mass, satisfy $r_{\rm ps}/r_{\rm AH}=3/2$. The diagram in Fig.~\ref{fig:ltb} shows the qualitative aspects of the LTB collapse model without a particular energy profile.

\subsection{Accreting/Evaporating Black Hole}

We now consider a dynamical scenario corresponding to a radiating black hole described by a solution with a radial flux of null dust. This setup represents both cases of ingoing or ongoing radiation to describe the energy gained or loss by a black hole. Physically, each case describes an accreting or evaporating black hole, respectively. A canonical example of this type of space-time is provided by the Vaidya metric  \cite{Vaidya:1951fdr}. 
Roughly speaking, the Vaidya metric describes a spherically symmetric spacetime containing a flux of null radiation and is generally regarded as one of the simplest dynamical extensions of the Schwarzschild solution, since it allows the black hole mass to evolve with time.

For this setup, we consider a basis on the time-radial manifold formed by the radial vector $n_a \equiv \nabla_a r/\sqrt{f}$ (given $f >0$), and a null, non-orthogonal vector $l_a$. This allows to study both accreting and evaporating scenarios at the same time\footnote{We exclude from our analysis those cases for which both processes happen simultaneously. Such cases need a double null basis to address them. An example of this is the description of mass inflation around the inner horizon of a black hole \cite{Poisson:1989zz}.}. The null vector can be defined through the Kodama-radial basis via
\begin{equation}
    l_a^{(\pm)} =  u_a \pm n_a,
\end{equation}
such that
\begin{align*}
    n_a n_b g^{ab} &= 1, \\ 
    l^{(\pm)}_a l^{(\pm)}_b g^{ab} &= 0, \\
    n_a l_b^{(\pm)} g^{ab} &= \pm 1.
\end{align*}

The source of the space-time is just the density of radiation, or radial flux, $\Phi \geq 0$. The corresponding energy-momentum tensor is written as
\begin{equation}
    T_{a b} =  \Phi l^{(\pm)}_a l^{(\pm)}_b, \quad P_\perp = 0,
\end{equation}
while on the other hand, using the Kodama and radial projection, we obtain the relations
\begin{align*}
    E &= 8 \pi T_{ab} u^a u^b = 8 \pi \Phi, \\
    P &= 8 \pi T_{ab} n^a n^b = 8 \pi \Phi,\\
    Q &= -8 \pi T_{ab} u^{(a} n^{b)}= \pm 8 \pi \Phi.
\end{align*}
which provides the energy, $E$, and, pressure, $P$, measured by the local Kodama and local radial observers. All these quantities are always positive. The radial heat, $Q$, shows that an evaporating black hole $Q>0$ uses the positive sign $(+)$ and the accreting black hole $Q<0$, the negative one ($-$).
Lastly, the gradient of the mass is

\begin{equation}
    \nabla_a m = \pm 4\pi r^2 \sqrt{f} \Phi l_a^{(\pm)}.
\end{equation}

It directly implies that the mass is constant along the direction of the radiation and changes for a local radial observer, that is
\begin{align}
    D_{l}^{(\pm)} m &= 0, \\
    D_n m &= 4\pi r^2 \sqrt{f} \Phi \ ,
\end{align}
respectively.

To analyze these scenarios in a specific coordinate system, we consider a generalized Vaidya metric in Eddington-Finkelstein coordinates $(w,r)$. Let $w$ be a null time coordinate defined through
\begin{equation}
    l_a^{(\pm)} = - \sqrt{f} \nabla_a w,
\end{equation}
which is the retarded time $u$ for outgoing radiation (evaporation) or the advanced time $v$ for ingoing radiation (accretion). The metric is:
\begin{equation}
    \der s^2 = -f(w,r) \der w^2 \mp 2 \der w \der r + r^2 \der\Omega^2,
\end{equation}
where the minus sign ($-$) corresponds to the evaporating case, and the plus sign ($+$) to the accreting case. The basis ($r_a$,$l_a^{(\pm)}$) in these coordinates is
\begin{align*}
    l_{a}^{(\pm)} &= \sqrt{f} \left( -1, 0\right) , & n_a &= \frac{1}{\sqrt{f}} \left(0, 1 \right), \\
    l^{a}_{(\pm)} &= \sqrt{f} \left( 0, \pm 1\right) , & n^a &= \frac{1}{\sqrt{f}} \left(\mp 1, f\right) \ ,
\end{align*}
respectively. Then, the derivative of the mass along the basis becomes
\begin{align}
D_{l}^{(\pm)}m&= \pm \sqrt{f} m' = 0, \\
    D_n m         &= \mp\frac{\dot{m}}{\sqrt{f}}.
\end{align}
This implies that the Misner-Sharp mass varies only on the null time, $m=m(w)$, and we express its rate of change as $\dot{m} \equiv \partial_w m$. Then, we derive the relation,

\begin{equation}
    \Phi = \mp \frac{\dot m}{4\pi r^2 f},
\end{equation}

which, implies that for evaporation the black hole loses mass, $\dot{m}<0$, while for accretion it gains energy, $\dot{m}>0$.

Substituting the above expressions into the general photon sphere equation \eqref{eq:ps_condition}, yields the coordinate-specific equation for the photon sphere location $r_\text{ps}$:
\begin{equation}
    1 - \frac{3m(w_\text{ps})}{r_\text{ps}} - \frac{\left| \dot{m}(w_\text{ps}) \right|}{f(w_\text{ps})}  = 0.
\end{equation}
Expanding the metric function $f = 1 - \frac{2m}{r}$, we arrive to
\begin{equation}
    \left( 1 - \frac{3m}{r}\right)\left( 1 - \frac{2m}{r}\right) - \left| \dot{m}(w_\text{ps}) \right| = 0,
\end{equation}
which is the condition obtained in \cite{Claudel:2000yi} for instantaneous limit in which the photon surface has no radial velocity. This is a quadratic equation for $r_\text{ps}$, namely
\begin{equation}
    \left[ ( 1 -  \left| \dot{m}\right| ) r^2 - 5 m r + 6 m^2 \right]_\text{ps} = 0, \label{eq:quadratic_ps2}
\end{equation}
that yields two solutions
\begin{equation}
    r_{\pm} = m (w_\text{ps}) \frac{ 5 \pm \sqrt{ 1 + 24 \left| \dot{m}(w_\text{ps}) \right| } }{2 \left( 1 - \left| \dot{m}(w_\text{ps}) \right| \right)}. \label{eq:quadratic_ps}
\end{equation}
However, when taking the Schwarzschild limit, $\dot m(w) = 0$, and $m(w) = M$, the solutions yield $r_{\pm} = \{2M,3M\}$. The first solution indicates a photon sphere on the trapping horizon, while the second is the standard location of the static photon sphere. Therefore, the only valid solution is the one with the positive sign, $r_+$, so we can write
\begin{equation}
    r_\text{ps}(w) = 3 m (w) \mathcal{A}[\dot m (w)], \label{eq:ps_vaidya}
\end{equation} 
with 
\begin{equation}
    \mathcal{A}[\dot m (w)] \equiv  \frac{ 5 +\sqrt{ 1 + 24 \left| \dot{m}(w) \right| } }{6 \left( 1 - \left| \dot{m}(w) \right| \right)}.
\end{equation}
The functional $\mathcal{A}[\dot m (w)]$ is always positive, given $0 \leq \left|\dot m \right|<1$, and is equal to one when there is no energy transfer $\left|\dot m \right|=0$.
At $\left|\dot m\right|=1$, the quadratic Equation~\eqref{eq:quadratic_ps} degenerates into the linear equation $-5mr+6m^2=0$, whose solution is $r=6m/5<2m$. The exterior branch $r_+$ diverges as $\left|\dot m\right|\to 1$ and does not provide a positive radius solution for $|\dot m|>1$. For $|\dot m|>1$, the remaining positive root is the continuation of $r_-$ and lies inside the trapping horizon. Consequently, no positive radius solution exists in the untrapped domain $f>0$ when $|\dot m|\geq1$.

In this adiabatic regime, where the mass changes slowly compared with the orbital time scale, we obtain
\begin{equation}
    \mathcal{A}[\dot m (w)] = 1+3\left|\dot m(w)\right|+\mathcal{O}(\left|\dot m\right|^2),
\end{equation}
This quasi-local radius shows similar results to coordinate-based dynamical photon surfaces prescriptions in the slow-evolution regime~\cite{Claudel:2000yi}.

The result in Eq.~\eqref{eq:ps_vaidya} gives the radii of a time-dependent family of instantaneous dynamical photon spheres in a generalized Vaidya space-time for a specified evaporating or accreting model $m(w)$. On any smooth admissible branch, the union of this family defines a photon sphere tube. This solution contains a divergence when $\left|\dot m\right| = 1$, where the quadratic equation \eqref{eq:quadratic_ps} turns linear with a unique solution, $r_\text{ps} = \frac{6}{5}m$, which is inside the trapping horizon as $r_\text{ps} < r_{\text{h}} = 2 m$. The outer solution diverges to spatial infinity in this case, effectively disappearing. Thus, $\left| \dot m \right| = 1$ is a critical energy transfer, where we lose a valid exterior critical-turning-point radius.
We can also characterize the local radial deviation near the photon sphere via the tidal curvature, which reads in this case as
\begin{equation}
    \mathcal{K}_\text{ps} = \frac{L^2}{r_\text{ps}^4} \left( \frac{6m(w_\text{ps})}{r_\text{ps}} - 1 \right).
\end{equation}
At a single Vaidya critical-turning event, $\mathcal{K}_{\rm ps}>0$ indicates a locally separating radial deviation between neighboring null geodesics. In the Schwarzschild limit, $m=M$ and $r_{\rm ps}=3M$, the reference orbit is a static circular null orbit and $\mathcal{K}_{\rm ps}=L^2/r_{\rm ps}^4>0$, recovering its standard instability.

Alternatively, one can use Eq. \eqref{eq:ps_vaidya} to find
\begin{equation}
    \mathcal{K}_\text{ps} = \frac{L^2}{r_\text{ps}^4} \left( \frac{2}{\mathcal{A}[\dot m (w)]} - 1 \right).
\end{equation}

For the exterior branch $0\leq|\dot m|<1$, this quantity depends on the instantaneous energy-transfer rate. At a critical turning point, the local radial deviation is separating when $r_{\rm ps}<6m$, restoring when $r_{\rm ps}>6m$, and marginal when $r_{\rm ps}=6m$. Since each instantaneous dynamical photon sphere on this branch also satisfies $r_{\rm ps}>2m$, the instantaneously separating regime is
\begin{equation}
    2m(w) < r_\text{ps}(w) < 6 m(w).
\end{equation}

Additionally, we can plug this condition in the quadratic equation~\eqref{eq:quadratic_ps2} to obtain the critical energy transfer:
\begin{equation}
    \left| \dot m \right|_\text{ps} = \frac{1}{3}.
\end{equation}
Within the domain and quasi-local interpretation, the regimes along the time-dependent family are:
\begin{itemize}
\item $0\leq|\dot m|_{\rm ps}<1/3$: the instantaneous dynamical photon sphere on the exterior branch satisfies $2m<r_{\rm ps}<6m$, and the local radial deviation between neighboring null geodesics is separating, $\mathcal{K}_{\rm ps}>0$.
\item $|\dot m|_{\rm ps}=1/3$: the instantaneous dynamical photon sphere satisfies $r_{\rm ps}=6m$ and the local radial deviation is marginal, $\mathcal{K}_{\rm ps}=0$.
\item $1/3<|\dot m|_{\rm ps}<1$: the instantaneous dynamical photon sphere on the exterior branch satisfies $r_{\rm ps}>6m$, and the local radial deviation between neighboring null geodesics is restoring, $\mathcal{K}_{\rm ps}<0$. This local sign does not by itself establish globally bounded orbital evolution.
\item $|\dot m|_{\rm ps}\geq1$: no positive-radius solution in the untrapped domain $f>0$ exists.
\end{itemize}

This interpretation differs from dynamical photon sphere studies in which the photon sphere is defined as a moving hypersurface generated by null geodesics and its trajectory is obtained from a differential evolution equation \cite{Claudel:2000yi,Mishra:2019trb,Solanki:2022glc}. Our result should be viewed as a quasi-local characterization of the local radial deviation at a critical turning point. We expect similar behavior in slowly evolving regimes, but the two notions are guaranteed to coincide only in the static limit, where the photon surface has constant areal radius.\\

For completeness, we treat some models that describe physical scenarios for evaporation and accretion cases.

\subsubsection{Linear evaporation/accretion model}
The simplest model considers a linear change in mass over time, where $\lambda$ is a constant rate of accretion or evaporation. For the exterior family considered below, $0\leq\lambda<1$.
\begin{equation}
    \dot m(w) = \pm \lambda, \quad m(w) = m_0 \pm \lambda w.    
\end{equation}
The photon sphere in this case is located at:
\begin{equation}
    r_\text{ps}(w) = 3 m(w) \mathcal{A}_\lambda,
\end{equation}
with
\begin{equation}
    \mathcal{A}_\lambda \equiv \frac{5 + \sqrt{1 + 24\lambda}}{6(1 - \lambda)},
\end{equation}
a constant that depends only on the rate parameter $\lambda$. Then, the location of the photon sphere relative to the trapping horizon is a constant,
\begin{equation}
    \frac{r_\text{ps}}{r_\text{h}} = \frac{3}{2} \mathcal{A}_\lambda,
\end{equation}
Thus, the radius in this time-dependent family maintains a constant ratio to the trapping-horizon radius. The local radial deviation is also time independent and depends only on $\lambda$: it is separating for $\lambda<1/3$, marginal at $\lambda=1/3$, and restoring for $1/3<\lambda<1$. As $\lambda\to1^{-}$, the radius of the family diverges.

\subsubsection{Hawking evaporation}

The Hawking evaporation model describes a black hole losing mass due to quantum effects. The rate of evaporation is assumed to be inversely proportional to the square of its mass \cite{Hiscock:1980ze}:
\begin{equation}
    \dot m(w) = -\frac{\alpha}{m^2}, \quad m(w) = \left(m_0^3 - 3 \alpha w \right)^{\tfrac{1}{3}},
\end{equation}
with $\alpha >0$ a constant characterizing the radiation flux. The model reaches its evaporation endpoint when $m(w_\text{end})=0$, namely
\begin{equation*}
    w_\text{end} = \frac{m_0^3}{3 \alpha}.
\end{equation*}
The photon sphere is now located at
\begin{equation}
    r_\text{ps}(w) = 3 \left(m_0^3 - 3 \alpha w \right)^{\tfrac{1}{3}} \mathcal{A}(w),
\end{equation}
with
\begin{equation}
    \mathcal{A}(w) = \frac{5 + \sqrt{1 + 24 \frac{\alpha}{m(w)^2}} }{6 \left( 1 - \frac{\alpha}{m(w)^2} \right)} .
\end{equation}
Unlike the linear evaporating scenario, the Hawking model induces non-linear transitions on the photon sphere. As the black hole evaporates, the photon sphere passes through the following sequence:
\begin{enumerate}
    \item Initially, the black hole features a dynamical photon sphere located near the standard Schwarzschild photon sphere, $r_\text{ps}\simeq3m_0$, where the local radial deviation between neighboring null geodesics is separating. As evaporation proceeds, the photon sphere moves inward, tracking the shrinking trapping horizon.
    \item At some moment, the photon sphere stops collapsing and bounces back to expanding outwards, while the trapping horizon keeps collapsing. This bounce point is determined by the condition $\tfrac{\der r_\text{ps}}{\der w} = 0$, yielding:
    \begin{equation*}
        w_\text{bounce} = w_\text{end} - \frac{\sqrt{\alpha}}{3} \left(\frac{5 \sqrt{33} - 21}{2} \right)^{3/2}.
    \end{equation*}
    \item Some time after the bounce, the local radial deviation near the expanding photon sphere changes from separating to restoring, passing through the marginal value $\mathcal{K}_{\rm ps}=0$ when $|\dot m|=1/3$. The marginal event occurs at
    \begin{equation*}
        w_\text{marg} = w_\text{end}- \sqrt{3 \alpha}.
    \end{equation*}
    \item Before the evaporation ends at $w_\text{end}$, the expanding photon sphere accelerates until the radial coordinate diverges when $\left| \dot m \right| = 1$, at the time
    \begin{equation*}
        w_\text{div} = w_\text{end} - \frac{\sqrt{\alpha}}{3}.
    \end{equation*}
    Consequently, even when the radius of the photon sphere goes to zero at the end of the evaporation, $\displaystyle \lim_{w \to w_\text{end}} r_\text{ps} = 0$, the geometry of the photon sphere features a divergent phase before. Since no positive-radius solution exists after this event, the exterior branch cannot be continued to $w_{\rm end}$ and no physical limit of $r_{\rm ps}$ is assigned there.
\end{enumerate}

\subsubsection{Eddington accretion}

The Eddington accretion model describes the maximum rate at which a body can accrete matter when the outward radiation pressure balances the inward gravitational force \cite{Abramowicz:2011xu}. The mass grows exponentially with a rate $\gamma$, that is
\begin{equation}
    \dot m(w) = \gamma m, \quad m(w) = m_0 e^{\gamma w}.
\end{equation}
The photon sphere in this case is located at:
\begin{equation}
    r_\text{ps}(w) = 3 m_0 e^{\gamma w} \mathcal{A}(w),
\end{equation}
with
\begin{equation}
    \mathcal{A}(w) = \frac{ 5 + \sqrt{1 + 24\gamma m(w)}}{6(1 - \gamma m(w))}
\end{equation}
In this model, the photon sphere expands exponentially as the black hole accretes, following a sequence of phases defined by mass-change thresholds:
\begin{enumerate}
    \item Starting from an initial mass $m_0$, the photon-sphere radius increases from a value near the static Schwarzschild radius $3m_0$, while the local radial deviation between neighboring null geodesics is initially separating.
    \item The local radial deviation becomes marginal when the accreting mass hits $m=1/(3\gamma)$. At this moment, the exterior photon sphere is pushed to $r_{\text{ps}} = 2/\gamma$. This happens at a time given by
    \begin{equation*}
        w_\text{marg} = \frac{1}{\gamma} \ln \left( \frac{1}{3 \gamma m_0} \right).
    \end{equation*}

    \item If the accretion continues, the mass reaches the inverse of the exponential rate factor, $m(w)=1/\gamma$, where  the factor $\mathcal A(w)$ becomes singular. At this limit, the photon sphere diverges to infinity, at a time
    \begin{equation*}
    w_\text{div} = \frac{1}{\gamma} \ln \left( \frac{1}{\gamma m_0} \right).
    \end{equation*}
\end{enumerate}

This model shows that continuous mass addition initially produces an expanding family of instantaneous dynamical photon spheres with a locally separating radial deviation. As the accretion rate grows, the local radial deviation becomes marginal and then restoring, before the exterior critical turning point radius diverges at the limiting rate. These local classifications do not by themselves determine the global stability of neighboring null orbits.

\section{Conclusion} \label{sec:conclusion}

In this work we have introduced a covariant formalism to define and analyze dynamical photon spheres in dynamical, spherically symmetric space-times. Our approach is based on a 2+2 spherical space-time decomposition, and the use of the gradient of the areal radius, and its orthogonal, the Kodama vector, to define covariant quantities. 
In this way, we avoid relying on a time-like Killing symmetry and identify critical turning points $x_{\rm ps}\in\mathcal M_s$ through the instantaneous conditions $\Pi|_{x_{\rm ps}}=0$ and $\nabla_k\Pi|_{x_{\rm ps}}=0$. Each critical turning point determines a dynamical photon sphere $\mathcal S_{\rm ps}(x_{\rm ps})$; as the space-time evolves, these spheres form a time-dependent family whose smooth union defines the photon sphere tube. We derived a general algebraic equation for the location of the critical turning points that depends on the quasi-local quantities: the areal radius, the Misner-Sharp mass, and the local radial pressure of the source. 
Furthermore, using the tidal curvature, we established a condition for the local radial deviation between neighboring null geodesics at a single critical-turning event. This local classification is distinct from global stability, which requires evolving the geodesics or the deviation equation over a finite interval.

We have tested this framework with different examples. In particular, we have recovered the well known results for stationary Schwarzschild-like space-times like the Schwarzschild solution itself and the Reissner-Nordstr\"om one. In that limit, the photon sphere equation is expressed in terms of the expansion of the radial vector and, for an unstable static orbit, the tidal curvature is related to the square of its affine-parameter Lyapunov exponent. Such quantities make direct contact with those employed within the field of black hole imaging from accretion disks, and are also relevant for gravitational-wave phenomenology.

Next, we applied this framework to two dynamical scenarios of interest: a stellar collapse and a Vaidya-type space-time. In collapsing-star scenarios, such as the Oppenheimer--Snyder and Lema\^itre--Tolman--Bondi models, the critical-turning-point equation identifies a dynamical photon sphere in the stellar interior that can be present independently of a trapping horizon. As the collapsing models are pressureless, the exterior photon sphere coincides with the static photon sphere of a Schwarzschild solution, while the interior instantaneous photon sphere and trapping-horizon generally evolve differently during collapse.

As for the Vaidya space-time, it is typically employed in the literature to study radiating (evaporating/accreting) black holes. In this case, the instantaneous photon sphere radius and local radial deviation depend on the mass-transfer rate $\left|\dot m\right|$. We identified parameter values at which the local radial deviation changes character or the algebraic branches cease to yield a finite admissible radius. We furthermore explored these properties along the corresponding time-dependent families of instantaneous dynamical photon spheres for the simple linear evaporation/accretion models as well as for the more physically appealing Hawking and Eddington models of evaporation and accretion, respectively.

To summarize, in this work we have extended the static photon sphere definition to a quasi-local characterization of instantaneous dynamical photon spheres. We selected critical turning points of non-radial null motion rather than establishing circular trapping over a finite interval or defining a photon-surface hypersurface. Connecting this quasi-local characterization to dynamical space-time images or dynamical gravitational-wave signals requires global null-geodesic evolution and is left for future work. Furthermore, the framework may be applied to time-dependent compact configurations such as certain classes of boson stars, although determining their observational signatures likewise requires a global analysis. This work therefore provides a starting point for investigating how instantaneous dynamical photon spheres relate to the global propagation of radiation in evolving space-times.

\section*{Acknowledgments}

A.~R. would like to express his gratitude to Silesian University in Opava, Czech Republic for their financial support, and is very grateful for the hospitality of the University of Valencia (Spain), Valencia Polytechnic University (Spain) and
the Complutense University of Madrid (Spain). This work is supported by the Spanish National Grants PID2022-138607NB-I00 and CNS2024-154444, funded by  MICIU/AEI/10.13039/501100011033; and the grant program Vouchers for Universities in the Moravian-Silesian Region (registration number CZ.10.03.01/00/23\_042/00003901119). This article is based upon work from COST Action FuSe, CA24101, supported by COST (European Cooperation in Science and Technology).

\appendix
\section{Gravitational Field Equations} \label{sec:gravitational_field_equations}

Let us briefly work out the field equations associated to the $2+2$ decomposition of the manifold given by (\ref{eq:sph_metric_bg}). The non-vanishing components of the Ricci tensor for the full spherical space-time are given by (lowercase Latin indices correspond to the $x^a$ sector, while uppercase ones are referred to the angular sector)
\begin{subequations}
    \begin{align}
        ^{(4)} R_{ab} &= \frac{1}{2}\mathcal{R} g_{ab} - \frac{2}{r} r_{:ab},\\
        ^{(4)} R_{AB} &= \left(1 - r \square r - r_{:a} r^{:a} \right) \Omega_{AB},
    \end{align}
\end{subequations}
where $\square \equiv g^{ab}\nabla_{a} \nabla_{b}$ is the d'Alembert operator for time-radial coordinates, $\mathcal{R}_{ab}$ the curvature of the $\mathcal{M}_s$ manifold, and the curvature scalar is $\mathcal{R}\equiv g^{ab} \mathcal{R}_{ab}$. The Ricci scalar is
\begin{equation}
    ^{(4)}R = \mathcal{R} + \mathfrak{R},
\end{equation}
with $\mathfrak{R} \equiv \frac{2}{r^2} \left(1 - 2 r \square r - r_{:a} r^{:a}\right)$. Then, the Einstein tensor components read as
\begin{subequations}
\begin{eqnarray}
    ^{(4)}G_{ab} &=& - \frac{2}{r} r_{:ab} - \frac{\mathfrak{R}}{2} g_{ab}, \label{eq:einsteinBG1} \\
    ^{(4)}G_{AB} &=& \left( r \square r - \frac{1}{2}r^2 \mathcal{R}\right) \Omega_{AB} . \label{eq:einsteinBG2}
\end{eqnarray}\label{eq:einsteinBG}
\end{subequations}
Let us consider an Einstein tensor which is sourced by a spherically symmetric distribution of matter described by a source tensor as
\begin{equation}
    S_{\mu \nu} = 
    \begin{bmatrix}
        S_{ab} & 0 \\
        0 & P_\perp r^2 \Omega_{AB}
    \end{bmatrix},
\end{equation}
with $S_{ab}(x^a)$ the time-radial part of the stress tensor and $P_\perp(x^a)$ the elements on the diagonal for the 2-sphere. In such a case, the Einstein equations yield the following relations between the geometrical quantities and the generic source tensor under spherical symmetry, namely
\begin{align}
    \frac{r_{:ab}}{r}&= \frac{m}{r^3} g_{ab} - \frac{1}{2} \left( S_{ab} - S g_{ab} \right), \label{eq:r_ab} \\
    \mathcal{R} &= \frac{4m}{r^3} +  S - 2 P_\perp. \label{eq:EFE_2}
\end{align} 
with $S\equiv S_{ab}g^{ab}$ the contraction of $S_{ab}$.

\bibliographystyle{apsrev4-2}
\bibliography{references.bib}

\end{document}